\documentclass[showpacs,twocolumn]{revtex4-1}
\usepackage{amssymb}
\usepackage{graphicx}
\usepackage{subfigure}

\begin{document}

\title{Spatiotemporal vortex solitons in hexagonal arrays of waveguides}
\author{Herv\'{e} Leblond$^{1}$, Boris A. Malomed$^{2,3}$, and Dumitru
Mihalache$^{4,5,1}$}
\affiliation{$^{1}$Laboratoire de Photonique d'Angers, EA 4464, Universit\'{e} d'Angers, 2 Bd
Lavoisier, 49000 Angers, France\\
$^{2}$Department of Physical Electronics, School of Electrical
Engineering,
Faculty of Engineering, Tel Aviv University, Tel Aviv 69978, Israel\\
$^{3}$ICFO-Institut de Ciencies Fotoniques, Mediterranean Technology
Park,
08860 Castelldefels (Barcelona), Spain\footnote{temporary Sabbatical address}\\
$^{4}$Horia Hulubei National Institute for Physics and Nuclear
Engineering
(IFIN-HH), 407 Atomistilor, Magurele-Bucharest, 077125, Romania\\
$^{5}$Academy of Romanian Scientists, 54 Splaiul Independentei,
Bucharest 050094, Romania}

\begin{abstract}
By means of a systematic numerical analysis, we demonstrate that hexagonal
lattices of parallel linearly-coupled waveguides, with the intrinsic cubic
self-focusing nonlinearity, give rise to three species of stable
semi-discrete complexes (which are continuous in the longitudinal
direction), with embedded vorticity $S$: triangular modes with $S=1$,
hexagonal ones with $S=2$, both centered around an empty central core, and
compact triangles with $S=1$, which do not not include the empty site.
Collisions between stable triangular vortices are studied too. These
waveguiding lattices can be realized in optics and BEC.
\end{abstract}

\maketitle

\section{Introduction}

Lattice solitons are a topic of great interest to ongoing studies of
nonlinear dynamics in photonic media and BECs (Bose-Einstein condensates)
\cite{recent-reviews}. These localized modes are produced by the interplay
of the intrinsic nonlinearity of the medium with an effective periodic
potential induced in it by permanent or virtual lattice patterns. In fact,
the lattice may itself be a nonlinear structure if it is induced by a
spatially periodic modulation of the local nonlinearity \cite{BarcelonaRMP}.
In the limit of a deep periodic potential, the fundamental models of lattice
media reduce to various versions of the discrete nonlinear Schr\"{o}dinger
(DNLS) equation \cite{Panos}. The realization of the one-dimensional (1D)
DNLS model in arrayed optical waveguides was originally proposed in Ref.
\cite{Demetri}. The same model was later applied to BECs loaded into deep
optical-lattice potentials \cite{discrete} (see Ref. \cite{review-Chaos} for
a brief review). A physical realization of the DNLS model is also possible
in the form of lattices of microcavities which serve as traps for polaritons
\cite{polariton}.

Lattice solitons take the form of discrete solitons in terms of the DNLS
equations, which correspond to quasi-discrete solitons in the respective
experimental settings. Such solitons were created in a set of semiconductor
waveguides built on top of a slab substrate \cite{Silberberg}, and also in
arrays of optical fibers \cite{fiber-array}. In addition to using permanent
photonic structures, quasi-discrete solitons were also made in virtual
waveguiding arrays, using the versatile technique of inducing interference
lattices in photorefractive crystals \cite{Moti}. The latter method was used
to create the first examples of 2D quasi-discrete fundamental solitons \cite%
{Moti2D}, which was followed by the making of vortex solitons \cite{vortex},
i.e., localized lattice excitations with embedded vorticity, that were
predicted in Ref. \cite{vortex-prediction}. Another significant contribution
to this area was the creation of 2D solitons in a bundle of fiber-like
waveguides written in bulk silica \cite{JenaSoliton}. Such arrays and
bundles are created by means of tightly focused femtosecond laser pulses
\cite{JenaWriting}.

Following the analysis of the fundamental localized discrete vortices with
topological charge $S=1$ \cite{vortex-prediction}, their higher-order
counterparts, with $S>1$, and multipole discrete solitons, such as
quadrupoles, were predicted in Refs. \cite{S>1,HS}. Many other objects were
studied in this area, including \textit{supervortices} (circular chains of
compact vortices with an imprinted overall topological charge, which is
independent of the vorticity of the individual eddies \cite{HS}),
necklace-shaped patterns \cite{necklace}, discrete solitons in hexagonal and
honeycomb lattices \cite{hexa,experiment}, composite \textit{semidiscrete}
spatial solitons in arrays of waveguides with quadratic and cubic
nonlinearities \cite{Panoiu}, quasi-discrete topological solitons in
photonic-crystal fibers \cite{Ferrando}, etc. Nonstationary soliton effects
were studied too. These include the mobility of discrete solitons \cite%
{Boulder,Papa}, collisions between traveling ones \cite{Papa,interaction},
and the onset of the spatiotemporal collapse in self-focusing arrayed
waveguides \cite{Shimshon}.

Most works on lattice solitons dealt with the spatial-domain settings. In
particular, optically-induced lattices in photorefractive crystals do not
makes it possible to observe the evolution in the temporal domain because of
a very large response time in these materials. However, the spatiotemporal
dynamics can be realized in waveguiding arrays written in bulk silica \cite%
{JenaWriting}, where the spatially localized quasi-discrete patterns
in the transverse plane can be combined with the temporal
self-trapping in the longitudinal direction. Recently, the creation
of the corresponding quasi-discrete ``light bullets" was reported in
this system \cite{bullet} (for a review of spatiotemporal solitons
in nonlinear optics and BEC, see Ref. \cite{MMWT}). Previously, a
number of manifestations of the spatiotemporal self-trapping in
similar settings were studied theoretically, including the related
modulational instability \cite{MI}, formation of ``bullets" in fiber
arrays \cite{Turitsyn1} and photonic wires \cite{phot-wire}, and
self-compression \cite{compression} and steering \cite{steering} of
pulsed beams. Continuing the work in this direction, semi-discrete
spatiotemporal surface solitons were introduced, as
\textit{surface modes}, in models of semi-infinite waveguide arrays \cite%
{MihKiv}, and in a system with an interface between different arrays \cite%
{StaggUnstagg}. Also analyzed were spatiotemporal solitons in waveguide
arrays with the quadratic nonlinearity \cite{chi2array}.

Once stable fundamental spatiotemporal soliton complexes in bundled
arrays of waveguides are available, it is natural to seek for vortex
solitons in the same setting. A systematic analysis of
spatiotemporal vortices and quadrupoles in the model based on the
square lattice of discrete waveguides was reported in Ref.
\cite{we1}. A vast stability area was found for the solitary
vortices with $S~=1$ and quadrupoles, which are built as rhombuses,
alias on-site-centered modes, with respect to the underlying lattice
(the rhombus is built as a set of four ``bright" cores, with a
nearly ``dark" one at the center). The stability region is much
smaller for the off-site-centered modes of the ``square" type,
without an empty pivotal site in the middle (the reduced stability
domain of square-shaped vortices and quadrupoles, in comparison with
their rhombic counterparts, in a generic feature of topological
solitons in lattice media \cite{Thaw}). All the spatiotemporal
vortex solitons with $S~=2$ were found to be unstable unstable in
the same model. Further, collisions between stable vortices and
quadrupoles (with identical or opposite topological charges),
propagating along the bundle in opposite directions, were analyzed
in Ref. \cite{we2}. Four different outcomes of the collisions were
identified: rebound of slowly moving solitary vortices, fusion,
splitting, and quasi-elastic interactions between fast ones.

Hexagonal lattices may be created by means of the same techniques which were
used for the building the square-shaped structures. On the other hand, the
change of the underlying geometry may essentially alter fundamental
properties of topological lattice solitons \cite{Panos,hexa}. In particular,
it was predicted theoretically and conformed in an experiment that spatial
solitons in the form of double vortices (with $S=2$) in hexagonal lattices
may be stable, while their unitary counterparts (with $S=1$) are unstable.

The objective of the present work is to study spatiotemporal vortex solitons
in hexagonal lattices of discrete waveguides. The model is formulated in
Section 2, and at the end of it we also briefly consider fundamental
solitons, driven by a temporally self-trapped pulse in a single waveguiding
core. In Section 3, we demonstrate, also in a brief form, that a
straightforward input in the form of a hexagon-shaped spatiotemporal vortex
with $S=1$ always leads to a decay. Nevertheless, three different species of
\emph{stable} spatiotemporal complexes with the embedded vorticity are
revealed by a systematic numerical analysis. In Section 4, we demonstrate
that a spatiotemporal input of a triangular shape generates self-trapped
vortices in the form of triangles with an empty core in the middle. Further,
in Section 5 it is shown that a hexagonally shaped input with $S=2$ produces
stable spatiotemporal hexagons with the same (\emph{double}) topological
charge. Finally, a modified (shifted) input ansatz gives rise to stable
densely packed triangular vortices with $S=1$, without an empty central
core, as shown in Section 6. In addition to the study of these species of
spatiotemporal vortex solitons, in Section 7 collisions between
counterpropagating triangular ones are studied. The paper is concluded by
Section 8.

\section{The model and fundamental solitons}

\begin{figure}[tbp]
\includegraphics[width=8cm]{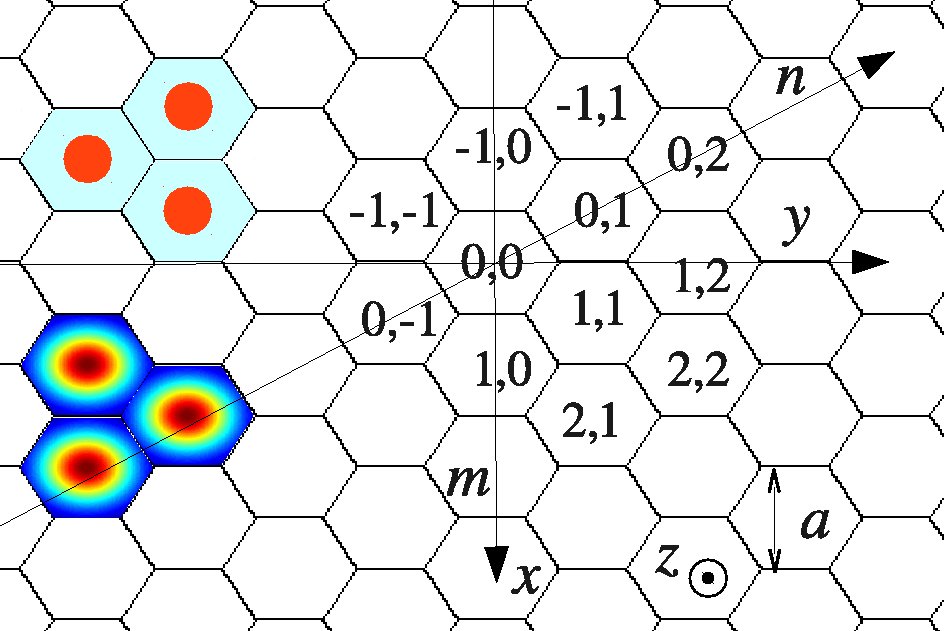}
\caption{(Color online) The setup and notation: We consider the hexagonal
array of cylindrical waveguides, as shown in the upper left corner. The
transverse distribution of the light intensity of the propagating waves 
in the guiding cores is displayed symbolically in the bottom
left corner. Each core is assumed to be a single-mode waveguide, represented
by wave function $u_{m,n}$, with discrete coordinates $\left( m,n\right) $
defined as shown in the figure. The map of integers $m,n$ into the Cartesian
coordinates in the transverse plane, $\left( x,y\right) $, is performed as
per Eq. (\protect\ref{xmn}), $a$ being the width of the hexagonal cell.}
\label{framec}
\end{figure}

We consider the hexagonal array of nonlinear waveguides, with cells in the
transverse lattice numbered as shown in Fig. \ref{framec}. The transmission
of waves in the array is described by the following system of coupled NLS
equations, written in the scaled form, similar to that used in many earlier
works \cite{MI}-\cite{we1}, \cite{bullet}:%
\begin{eqnarray}
i\partial _{z}u_{m,n} &+&\left[ u_{m-1,n-1}+u_{m-1,n}+u_{m,n-1}+u_{m,n+1}%
\right.  \nonumber \\
&&\left. +u_{m+1,n}+u_{m+1,n+1}-(6+\mu )u_{m,n}\right]  \nonumber \\
&+&(1/2)\partial _{t}^{2}u_{m,n}+u_{m,n}\left\vert u_{m,n}\right\vert ^{2}=0.
\label{eq1}
\end{eqnarray}%
In terms of the optical setting, $z$ and $t$ are, respectively, the
propagation distance and reduced time, assuming that each guiding core
features the anomalous chromatic dispersion and cubic self-focusing, while $%
-\mu $ is the propagation constant of the localized solution to be sought
for. In terms of the corresponding BEC model, Eqs. (\ref{eq1}) is a system
of coupled discrete Gross-Pitaevskii equations \cite{GP}, with $z$ and $t$
playing the roles of the scaled time and axial coordinate, respectively,
while $\mu $ is the chemical potential.

Simulations of Eq. (\ref{eq1}) were carried out in the Fourier domain, with
the help of the standard fourth-order Runge-Kutta scheme, the nonlinear term
being evaluated by means of the combination of inverse and direct fast
Fourier transforms at each sub-step of the scheme. We used an $11\times 12$
matrix in the plane of $\left( m,n\right) $, 512 points for variable $t$ in
the computation window of width $\Delta t=20$, and the stepsize in the
propagation direction $dz=5\times 10^{-4}$. The use of the Fourier transform
implies periodic boundary conditions in $t$, which make sense if a
characteristic temporal size of the localized objects will be essentially
smaller than $\Delta t=20$. As concerns the boundary conditions for the
discrete coordinates $m$ and $n$, the values of $u_{m,n}$ corresponding to
the coordinates which fall outside of the computation box are replaced by
zeros. 

Before proceeding to the search for complex spatiotemporal vortical
patterns, it makes sense to test the propagation of fundamental solitons,
which are carried, essentially, by a temporal pulse in a single core. For
this purpose, the simulations were initiated with obvious initial
conditions,
\begin{equation}
u_{0,0}(z=0)=\eta ~\mathrm{sech}\left( \eta t\right) ,  \label{fond}
\end{equation}%
\begin{equation}
\eta =\sqrt{2\left( 6+\mu \right) },  \label{eta}
\end{equation}%
setting $u_{m,n}(z=0)=0~\mathrm{at}~\left\vert m\right\vert +\left\vert
n\right\vert \neq 0$. The simulations were run in interval $1\leq \mu \leq
16 $ of values of the propagation constant.

\begin{figure}[tbp]
\subfigure[]{\includegraphics[width=8cm]{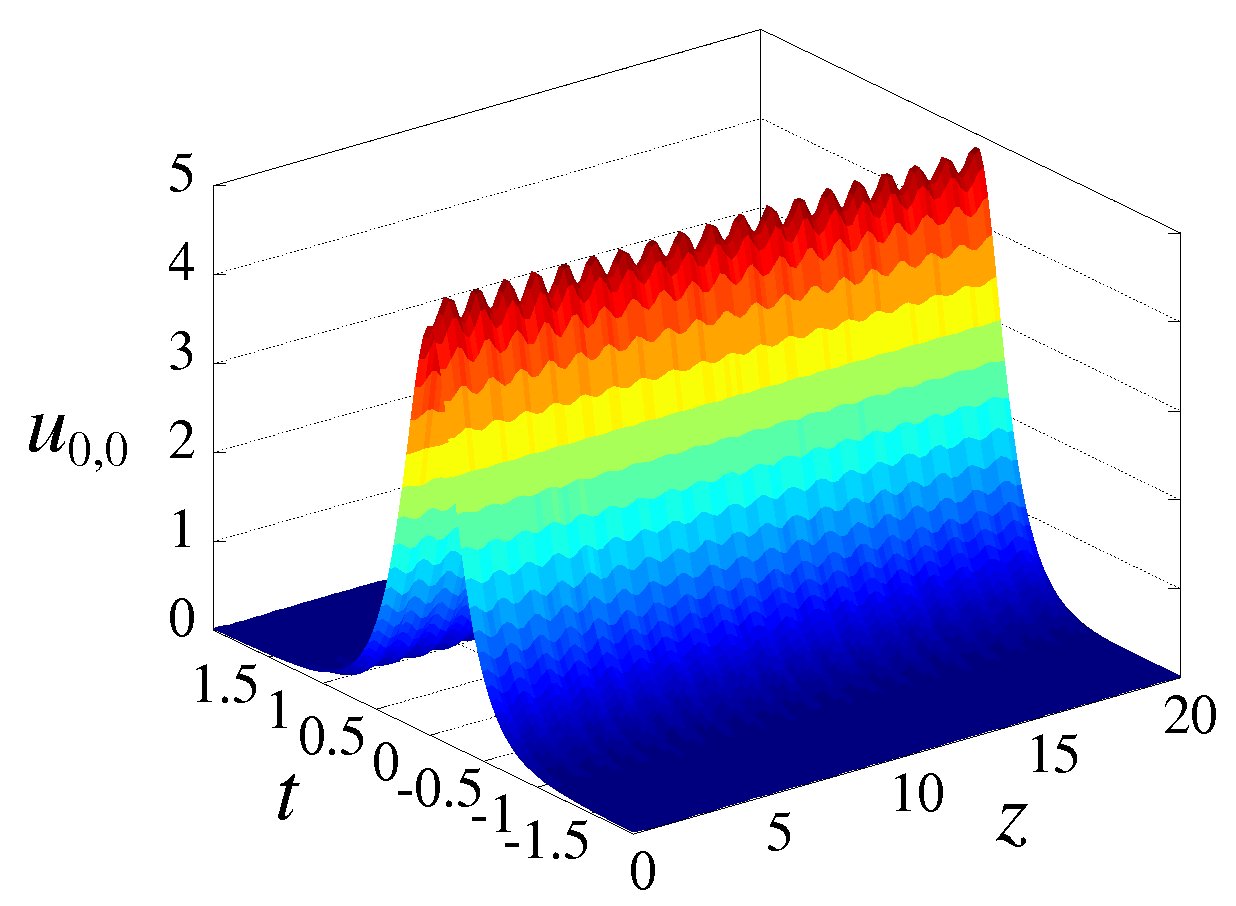}} \subfigure[]{%
\includegraphics[width=8cm]{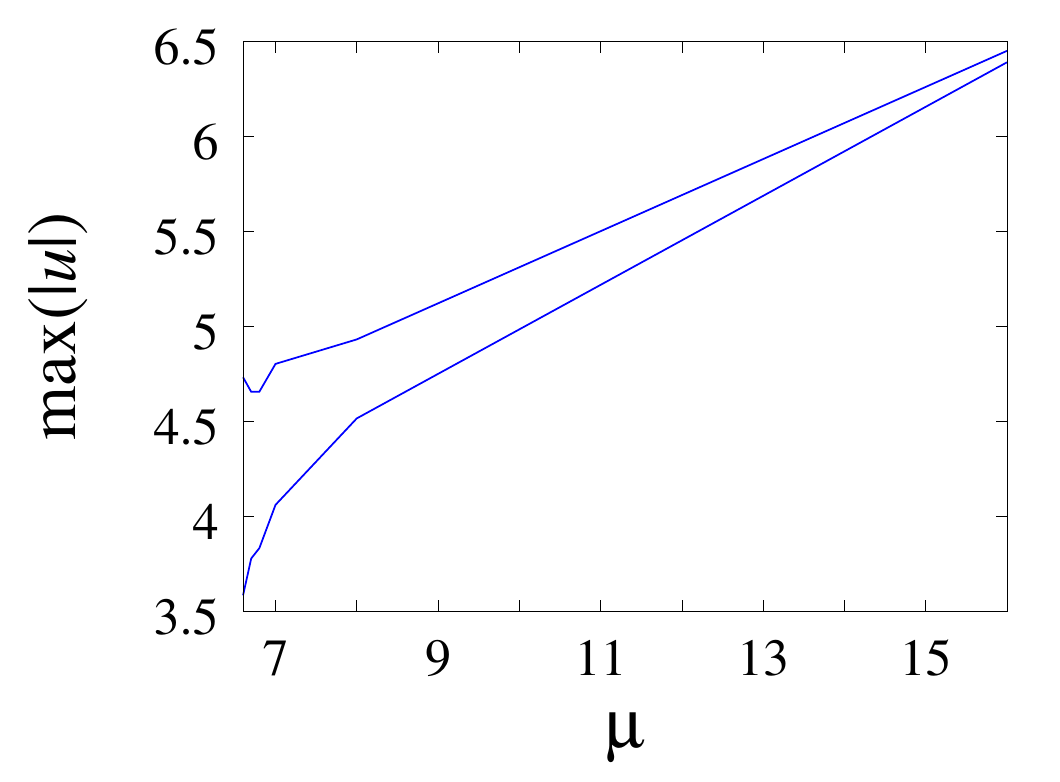}}
\caption{(Color online) (a) Oscillations of the fundamental soliton
generated by input (\protect\ref{fond}), (\protect\ref{eta}) with $\protect%
\mu =8$. (b) The upper and lower curves illustrate the oscillations by
showing, respectively, the largest and smallest values of the soliton's
amplitude, i.e., $\max_{z}\left( \max_{t}\left( |u|\right) \right) $ and $%
\min_{z}\left( \max_{t}\left( |u|\right) \right) $, as functions of the
propagation constant, $\protect\mu $.}
\label{osc_fond}
\end{figure}

It has been concluded that input (\ref{fond}) decays, under the action of
the lattice diffraction, at $\mu <6.6$, and a stable fundamental soliton,
concentrated in the central core, is formed at $\mu \geq 6.6$. The temporal
pulse which lies at the core of the so created fundamental soliton is not
quite stationary, but rather features regular pulsations, as shown in Fig. %
\ref{osc_fond}.

One may surmise that the oscillations of the fundamental soliton
could be a result of its interaction with the radiation background,
which was generated by the input field in the course of
self-trapping into the fundamental soliton. To check this
possibility, the background around the soliton was explicitly
removed, at a particular step of the simulations. Nevertheless, the
oscillations remain virtually unaffected by the ``cleaning", i.e.,
they seem to be a genuine feature of the dynamics of the soliton,
possibly representing its intrinsic mode.

\section{The hexagonal input: a transition to instability}

First, we attempted to create hexagonal vortical modes with $S=1$, which
seems a natural approach to the system based on a hexagonal lattice. To this
end, we used the following input, based on an ansatz factorized in the
longitudinal (temporal) and transverse (spatial) directions, cf. Ref. \cite%
{Japan}:%
\begin{eqnarray}
u_{m,n} &=&\frac{\left( x_{m,n}+iy_{m,n}\right) }{a}\exp \left[ -\alpha
\left( \sqrt{x_{m,n}^{2}+y_{m,n}^{2}}-a\right) \right]  \nonumber \\
&&\times \eta ~\mathrm{sech}\left( \eta t\right) ,  \label{hexag}
\end{eqnarray}%
with $\eta $ taken as per Eq. (\ref{eta}), and%
\begin{equation}
x_{m,n}\equiv a\left( m-\frac{n}{2}\right) ,~y_{m,n}\equiv \frac{\sqrt{3}}{2}%
an,  \label{xmn}
\end{equation}%
\begin{equation}
\alpha =\ln \left( 2\left( 6+\mu \right) \right) ,  \label{a}
\end{equation}%
$a$ being the width of the hexagonal cell, see Fig. \ref{framec}. The
model's scale is fixed by setting $a\equiv 1$. Factor $\left(
x_{m,n}+iy_{m,n}\right) $ in Eq. (\ref{hexag}) obviously corresponds to
vorticity $S=1$, and the exponential factor with $\alpha $ taken as per Eq. (%
\ref{a}) is determined as in the 2D spatial soliton with propagation
constant $-\mu $. The choice of $\eta $ as per Eq. (\ref{eta}) implies that,
simultaneously, the wave field in the factorized ansatz is localized in the
longitudinal direction, in each core, as in the temporal soliton
corresponding to the same propagation constant, cf. the structure of the
fundamental solitons considered above. The phase and energy structure of
ansatz (\ref{hexag}) is illustrated, in a schematic form, by Fig. \ref%
{hexag_trans}.
\begin{figure}[tbp]
\includegraphics[width=8cm]{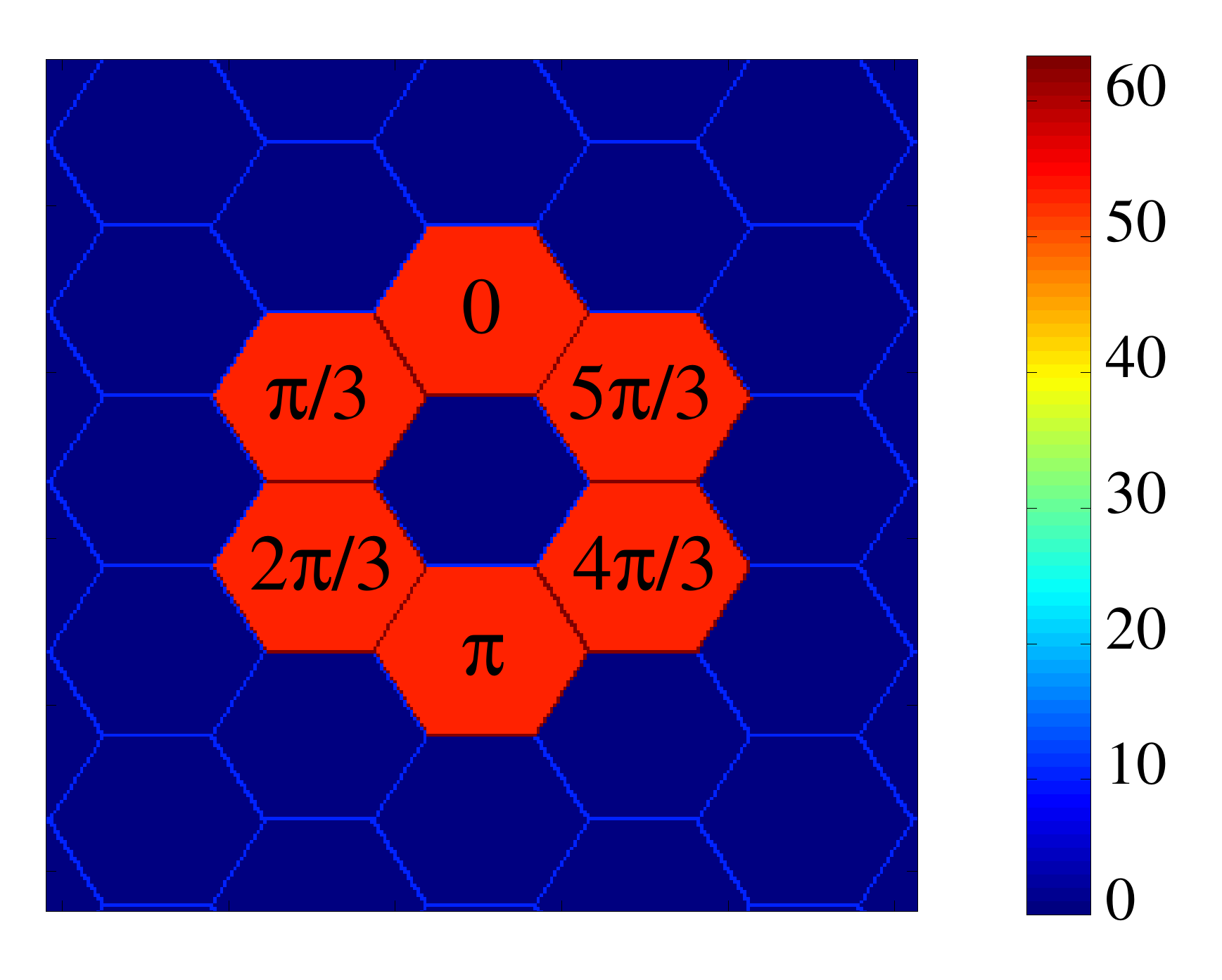}
\caption{(Color online) The phase and energy 
($\int_{-\infty}^{+\infty}|u_{m,n}|^{2}dt$) patterns
corresponding to input ansatz (\protect\ref{hexag}) which generates an
(unstable) hexagonal vortex with $\protect\mu =350$. }
\label{hexag_trans}
\end{figure}

\begin{figure}[tbp]
\includegraphics[width=8cm]{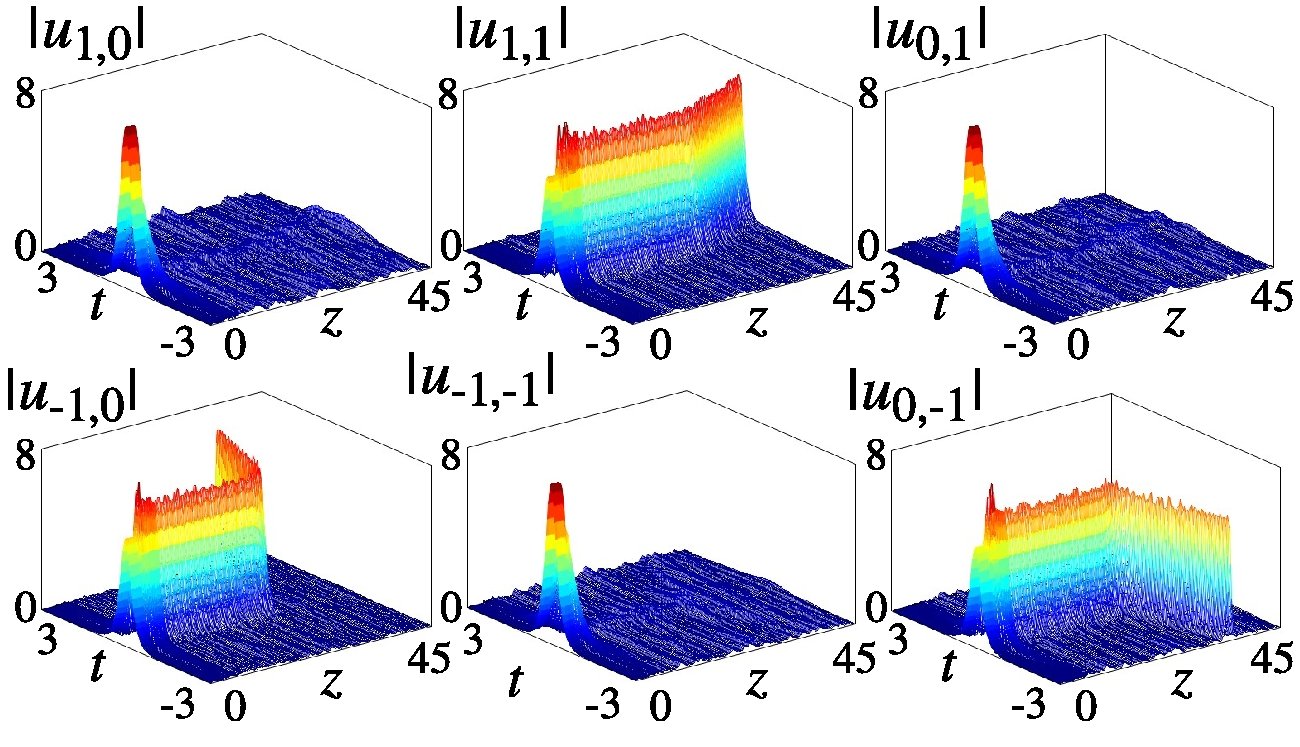}
\caption{(Color online) The evolution of the six main components of the
hexagon with initial propagation constant $\protect\mu =9$. In this case,
the simulations demonstrate the decay of the hexagon into a transient
triangle, which is followed by a longitudinal instability (splitting). }
\label{hexag_neuf}
\end{figure}

Direct simulations of Eq. (\ref{eq1}) with this input have been run in a
broad range of values of the propagation constant, $7<\mu <500$.
Nonetheless, stable spatiotemporal vortices with the hexagonal structure and
topological charge $S=1$ have never emerged. In fact, the evolution of the
input organized as the ansatz of this type never leads to formation of any
stable pattern. In the interval of $7\leq \mu \leq 13$, the system makes an
attempt to generate a robust pattern of a triangular shape, as shown in Fig. %
\ref{hexag_neuf}: at three sites belonging to the original hexagon, the
field quickly decays, while at three others it survives, for a while.
However, the largest amplitude of the temporarily emerging triangular set is
$\simeq 8$ (it is attained at $\mu =13$), while triangular vortices may be
stable for amplitudes above a threshold value of the amplitude which is $%
\simeq 18$ (see below), therefore the triangles developing from the unstable
hexagons are also subject to an instability, eventually splitting into
uncorrelated single-core excitations which separate in the longitudinal
direction, see Fig. \ref{hexag_neuf}.

Further, in the interval of $13<\mu \leq 70$, the instability splits the
original hexagon into a set of separating single-core excitations, the
number of which varies randomly between $2$ and $6$ (not shown here in
detail). In an adjacent interval, $100\leq \mu <200$, the
instability-development scenario is similar but faster, so that the
formation of a transient triangular structure cannot be identified.

At largest values of the propagation constant, $200\leq \mu \leq 500$, the
instability-development scenario is different. The six sites forming the
hexagon keep their positions and amplitudes for a while, but loose the
mutual phase coherence. Then, instabilities of amplitudes and positions set
in, but they manifest themselves on a much longer scale of the propagation
distance, with $z$ ranging from $10$ to a few hundreds, instead of $z\sim 1$
at small $\mu $, cf. Fig. \ref{hexag_neuf}. The separation between
excitations in individual cores grows very slowly too, in comparison with
the quick split of the transient triangle observed in Fig. \ref{hexag_neuf}%
.

\section{The generation of stable triangular vortices}

\begin{figure}[tbp]
\includegraphics[width=6cm]{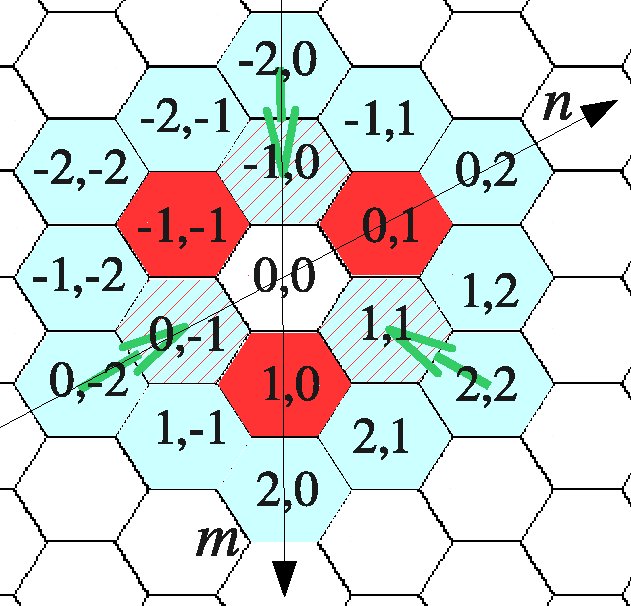}
\caption{(Color online) The reduction of the hexagonal input to the
triangular-vortex one: three of the six main peaks are replaced by fields
taken from the surrounding layer, as shown by arrows (the replacement makes
the amplitudes at the corresponding sites much smaller, but does not alter
their phases).}
\label{triangle}
\end{figure}

The next step is an attempt to generate a triangular vortical structure,
which is suggested by the emergence of a transient one in the course of the
evolution of the unstable hexagon (Fig. \ref{hexag_neuf}). For this purpose,
we used the same input as defined by Eqs. (\ref{hexag})-(\ref{a}), but with
three main peaks suppressed, which was done by replacing the fields at the
corresponding sites by those from adjacent sites in the outer layer: $%
u_{-1,0}\rightarrow u_{-2,0}$, $u_{1,1}\rightarrow u_{2,2}$, $%
u_{0,-1}\rightarrow u_{0,-2}$, as shown in Fig. \ref{triangle}. The so
constructed triangular ansatz keeps the vorticity of the original hexagon, $%
S=1$.

The evolution of this input was simulated in a broad range of values of the
propagation constant, $4\leq \mu \leq 450$. At small values, $\mu \leq 6,$
the three main peaks forming the triangle merge into a single-core
fundamental soliton, which may be localized at the central site, or at any
one belonging to the original triangle. Thus, stable vortices do not emerge
in this case. For $7\leq \mu \leq 181$, the triangle is destroyed by an
instability which splits it into separating uncorrelated single-core
excitations (the instability develops faster with the increase of $\mu $).

\begin{figure}[tbp]
\subfigure[]{\includegraphics[width=6cm]{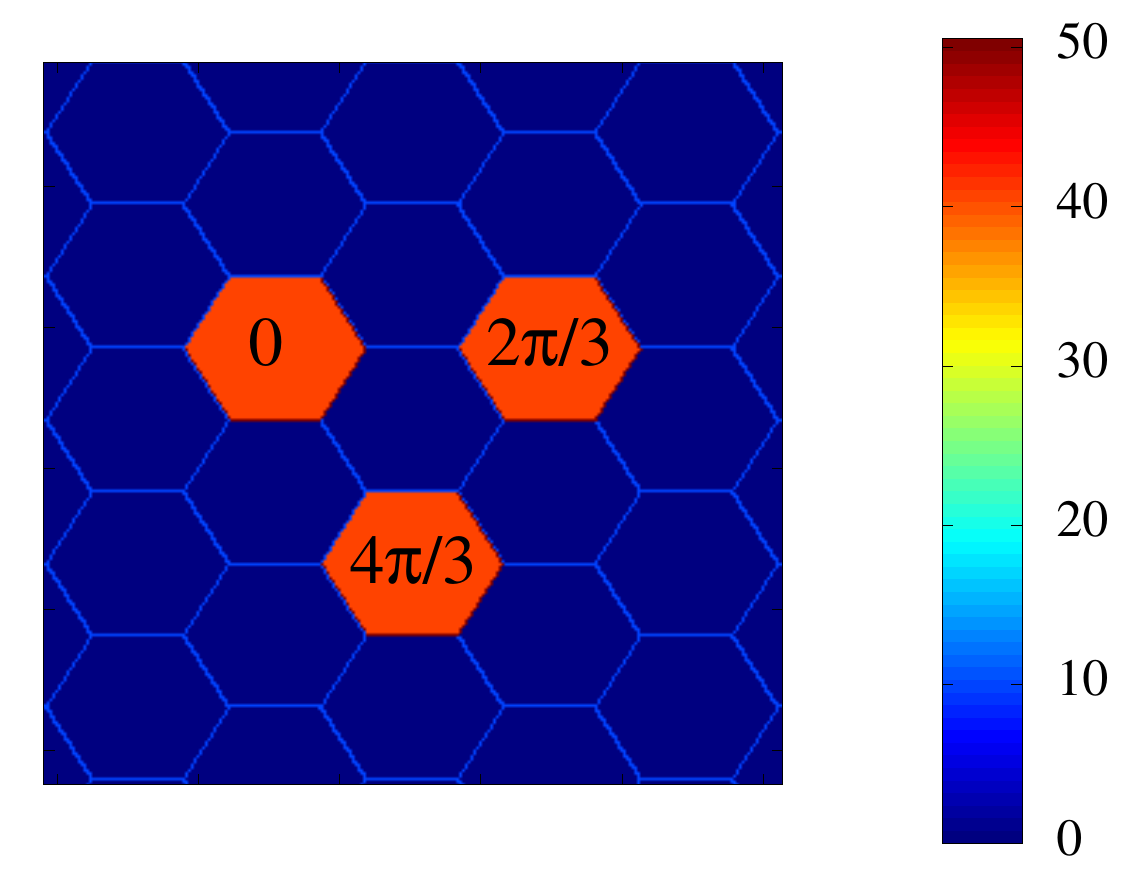}} %
\subfigure[]{\includegraphics[width=6cm]{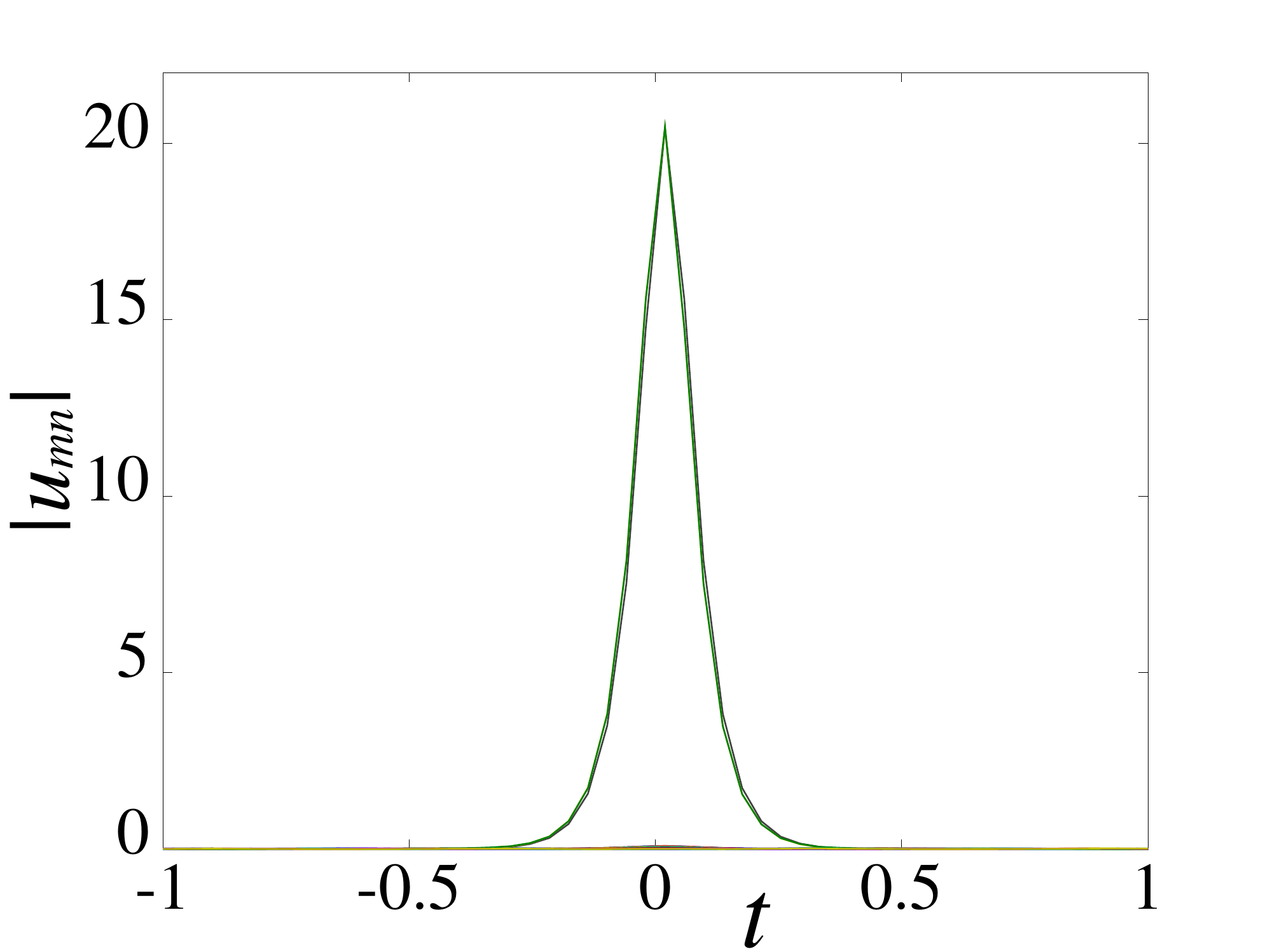}}
\caption{(Color online) (a) The transverse energy ($\int_{-\infty}^{+\infty}|u_{m,n}|^{2}dt$)
and phase
profile of the \emph{stable} triangular vortex generated by the input with $%
\protect\mu =200$. (b) The longitudinal (temporal) profile of excitations in
the cores representing the vertices of the triangle.}
\label{triangle_200}
\end{figure}

\begin{figure}[tbp]
\includegraphics[width=4cm]{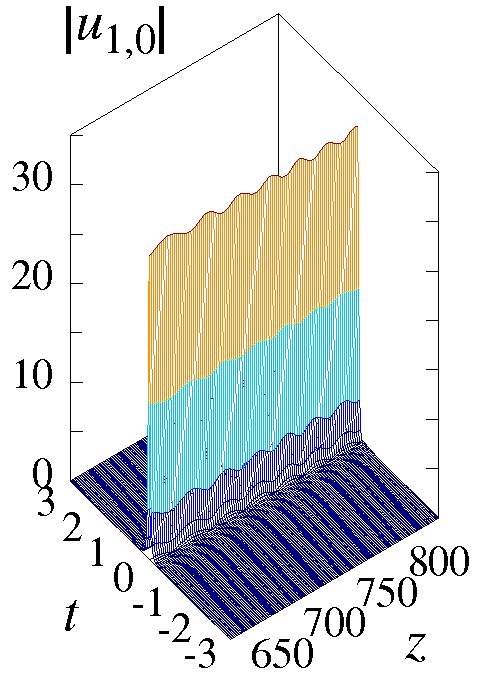} %
\includegraphics[width=4cm]{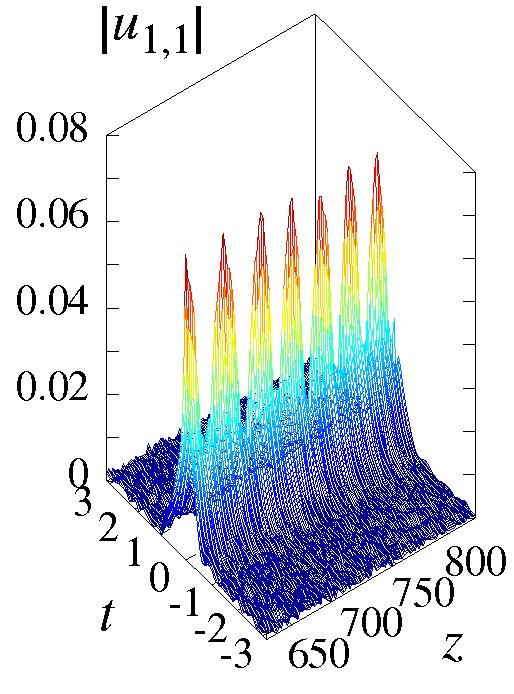}
\caption{(Color online) The evolution of the longitudinal (temporal)
profiles of one primary and one secondary components of a stable triangular
spatiotemporal vortex, generated by the triangular input with $\protect\mu %
=450$.}
\label{triangle_mumax}
\end{figure}

Finally, the same input generates \emph{stable} triangular vortices at $\mu
\geq 182$, an example of which is shown in Fig. \ref{triangle_200}. The
stability was verified by direct simulations for long propagation distances,
e.g., $z=986$ for $\mu =182$.
The temporal pulses in the cores which represent vertices of the triangle
remain well phase-locked, keeping the phase circulation of $2\pi $, which
corresponds to vorticity $S=1$.
The excitations at secondary sites (between the vertices) are phase-locked
to the primary ones, but featuring some oscillations. The oscillations
enhance with $\mu $, but the overall vortical phase pattern always persists.
On the other hand, the amplitudes of excitations at the secondary sites
feature fast irregular oscillations, which also become stronger at larger $%
\mu $ (variations of these amplitudes by a factor $\sim 2$ are observed
already at the stability threshold, $\mu =182$). These amplitude
oscillations are coupled to small variations of amplitudes at the primary
sites, as shown in Fig. \ref{triangle_mumax}. It has been checked that the
oscillations were not induced by reflection of small-amplitude radiation
waves from edges of the integration domain (absorbers installed at the edges
do not suppress the oscillations).

\begin{figure}[tbp]
\includegraphics[width=3.9cm]{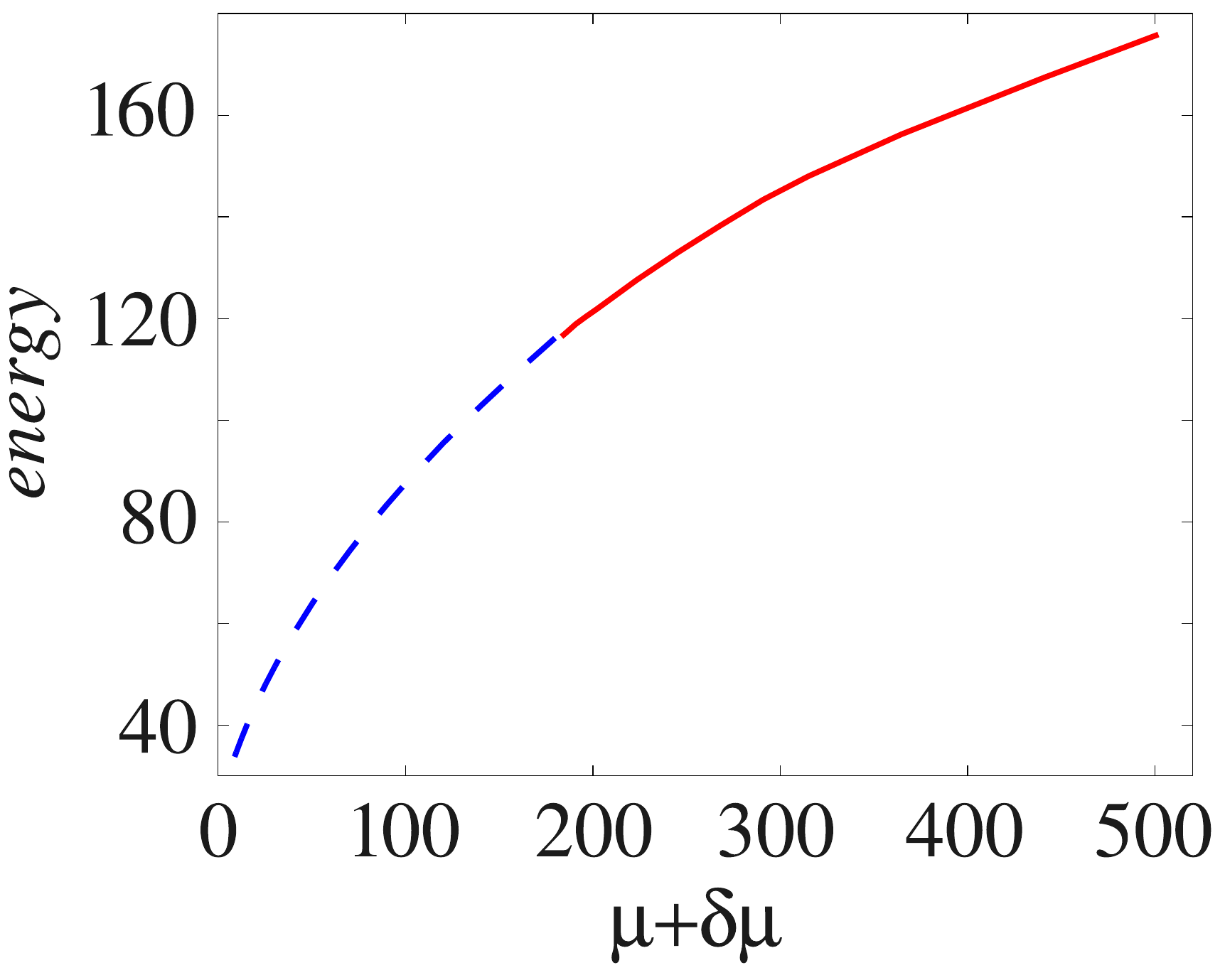} %
\includegraphics[width=3.9cm]{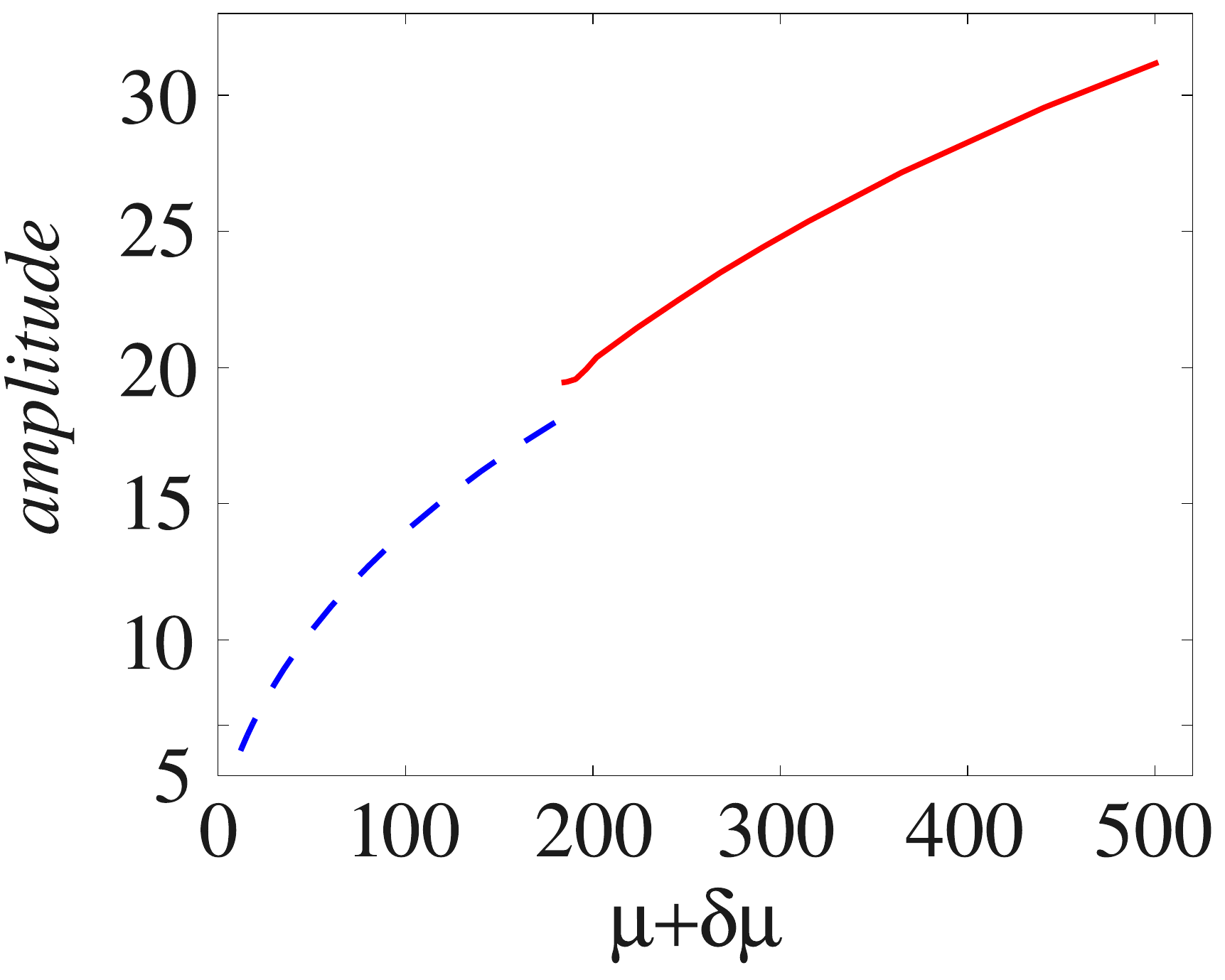}
\caption{(Color online) The amplitude and energy of the stable triangular
vortex vs. the effective propagation constant, $\protect\mu +\protect\delta
\protect\mu $, see Eq. (\protect\ref{deltamu}). Blue (dashed) and red (solid) segments
designate unstable and stable solution families, respectively.}
\label{triangle_curve}
\end{figure}

Figure \ref{triangle_curve} displays the total energy, $E=\sum_{m,n}\int_{-%
\infty }^{+\infty }\left\vert u_{m,n}(t)\right\vert ^{2}dt$, and amplitude
of the triangular vortices, both stable and unstable ones, as functions of
the effective propagation constant, $\mu +\delta \mu ,$ where the
contribution $\delta \mu $ from oscillations of the fields is computed as
follows. For each vertex of the triangle, $(m,n)$, peak time $t_{m,n}$ is
defined, such that $\left\vert u_{m,n}\left( t_{m,n}\right) \right\vert
=\max_{t}\left\vert u_{m,n}\left( t\right) \right\vert ,$ and the
corresponding phase, $\phi _{m,n}(t_{m,n})$, is identified. Next, we compute
\begin{equation}
\delta \mu =\left\langle \frac{d}{dz}\phi _{m,n}\right\rangle ,
\label{deltamu}
\end{equation}%
where the average is taken over the three vertices of the triangle (or six
ones for stable hexagonal vortices with $S=2$, see the next section), and,
for the stable triangular modes, over $z$. For unstable triangles, the
latter average was taken over a short interval $\Delta z$, within which the
pattern was not disturbed by the instability. As concerns the amplitude
shown in Fig. \ref{triangle_curve}, it was defined as $\left\vert
u_{m,n}\left( t_{m,n}\right) \right\vert $, averaged over $z$ and over the
three vertices, to smooth effects of small persistent oscillations of the
local amplitudes.
A small gap between the unstable and stable portions of the amplitude plot
in Fig. (\ref{deltamu}) is due to the 
difference in wavenumber shift (\ref{deltamu}), as computed for the stable
and unstable solutions.

\section{Hexagonal vortices with the double topological charge}

\begin{figure}[tbp]
\subfigure[]{\includegraphics[width=6cm]{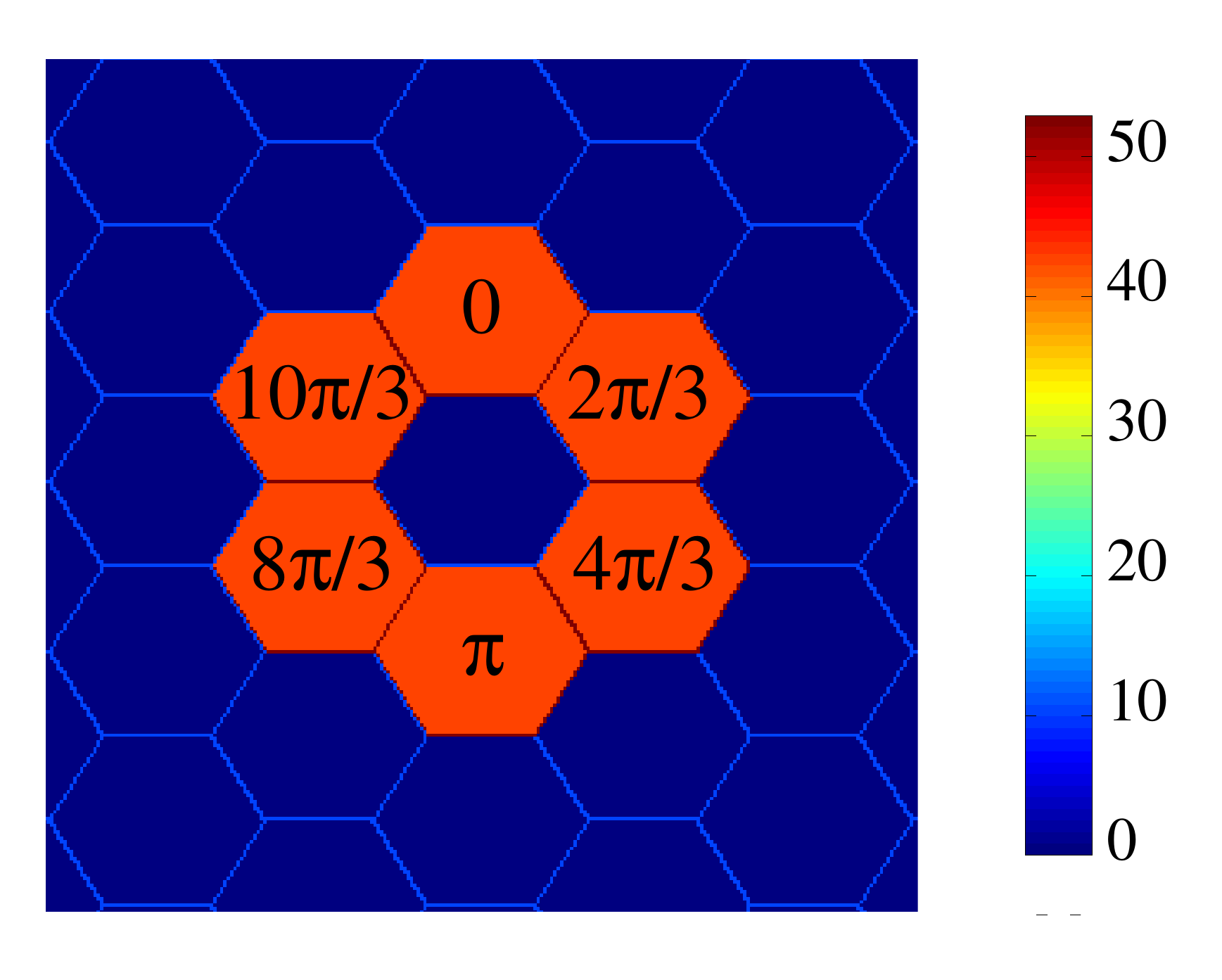}} %
\subfigure[]{\includegraphics[width=6cm]{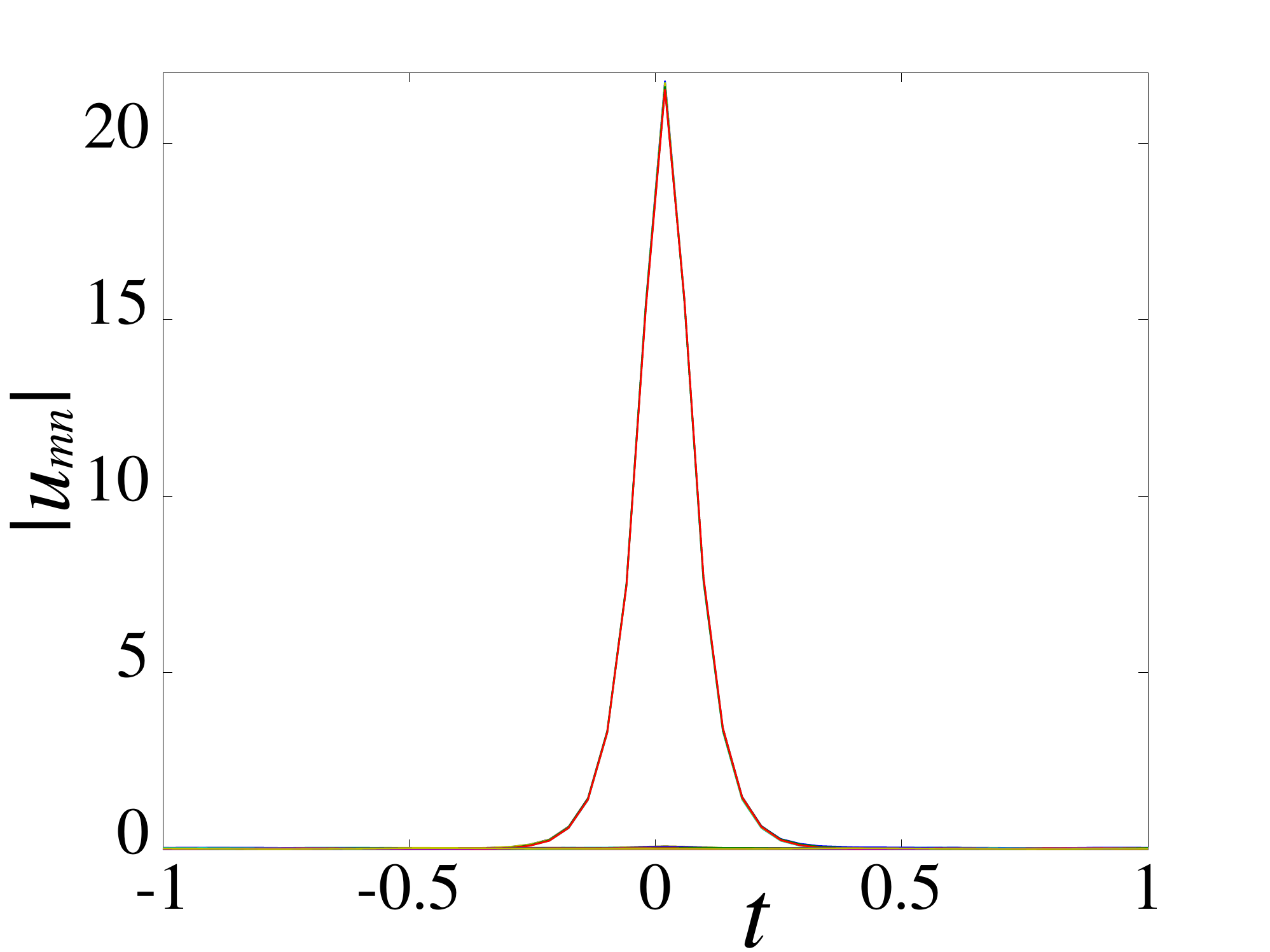}}
\caption{(Color online) The same as in Fig. \protect\ref{triangle_200}, but
for a stable hexagonal vortex with topological charge $S=2$, generated by
the modified input with $\protect\mu =223$.}
\label{double_223}
\end{figure}

 As said above, the
hexagonal input based on Eqs. (\ref{hexag})-(\ref{a}) could not
produce any stable pattern. However, the same initial ansatz, but
with inverse signs of three of its main peaks---say $u_{-1,0}$,
$u_{1,1}$, $u_{0,-1}$---can give
rise to \emph{stable} hexagonal spatiotemporal patterns carrying vorticity $%
S=-2$ (if the topological charge of the original ansatz is defined as $S=+1$%
), see an example of the stable mode in Fig. \ref{double_223}. Note that no
changes were made to the hexagonal input at sites in outer layers.

Simulations with this input were also run in a broad range of values of the
propagation constant, $5\leq \mu \leq 450$. At $\mu <223$, the seeded
pattern is subject to various instabilities: spread out due to the lattice
diffraction and temporal dispersion, or splitting into separating
excitations, or, sometimes, merger into one or two single-core fundamental
solitons. In particular, in the interval of $8\leq \mu \leq 10$, the pattern
forms a transient triangular structure, which eventually splits, and in a
broad interval of $20\leq \mu \leq 222$ the initial hexagon fissions into
two triangles, which also turn out to be unstable---essentially, because the
amplitudes of the triangular patterns fall below the stability threshold.

\emph{Stable} double ($S=2$) hexagonal vortices emerge at $\mu \geq 223$,
whose shape is illustrated by Fig. \ref{double_223}. The stability of these
vortices was verified in very long simulations. With the further increase of
$\mu $, an instability island was revealed around $\mu =300$. In that case,
the six temporal pulses remain locked to their positions, but the phase
structure is lost at $z\gtrsim 500$. Nevertheless, the hexagonal vortices
recover their integrity at still larger $\mu $. It is possible that other
narrow intervals of the instability may be found inside the stability region.

\begin{figure}[tbp]
\includegraphics[width=3.9cm]{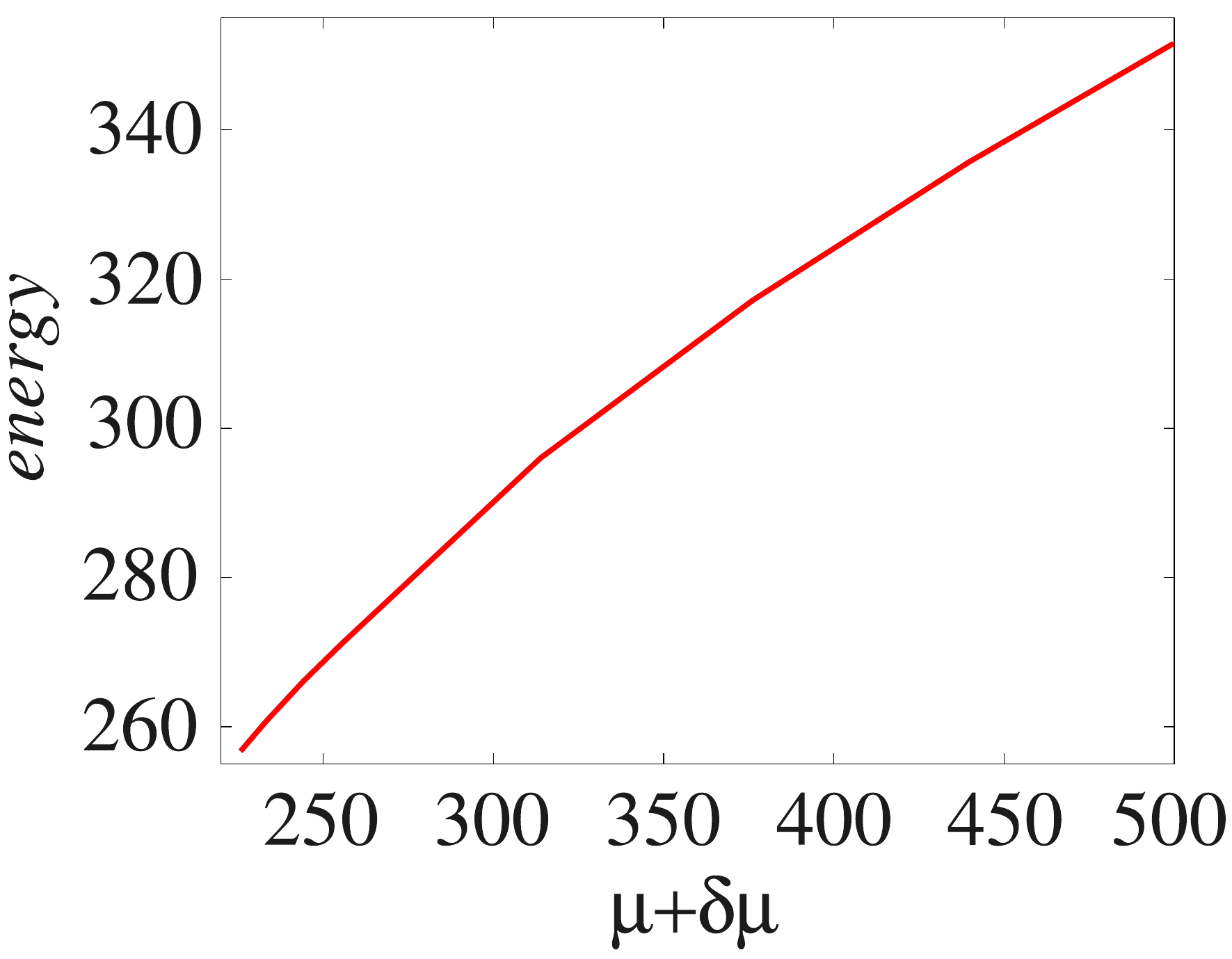} %
\includegraphics[width=3.9cm]{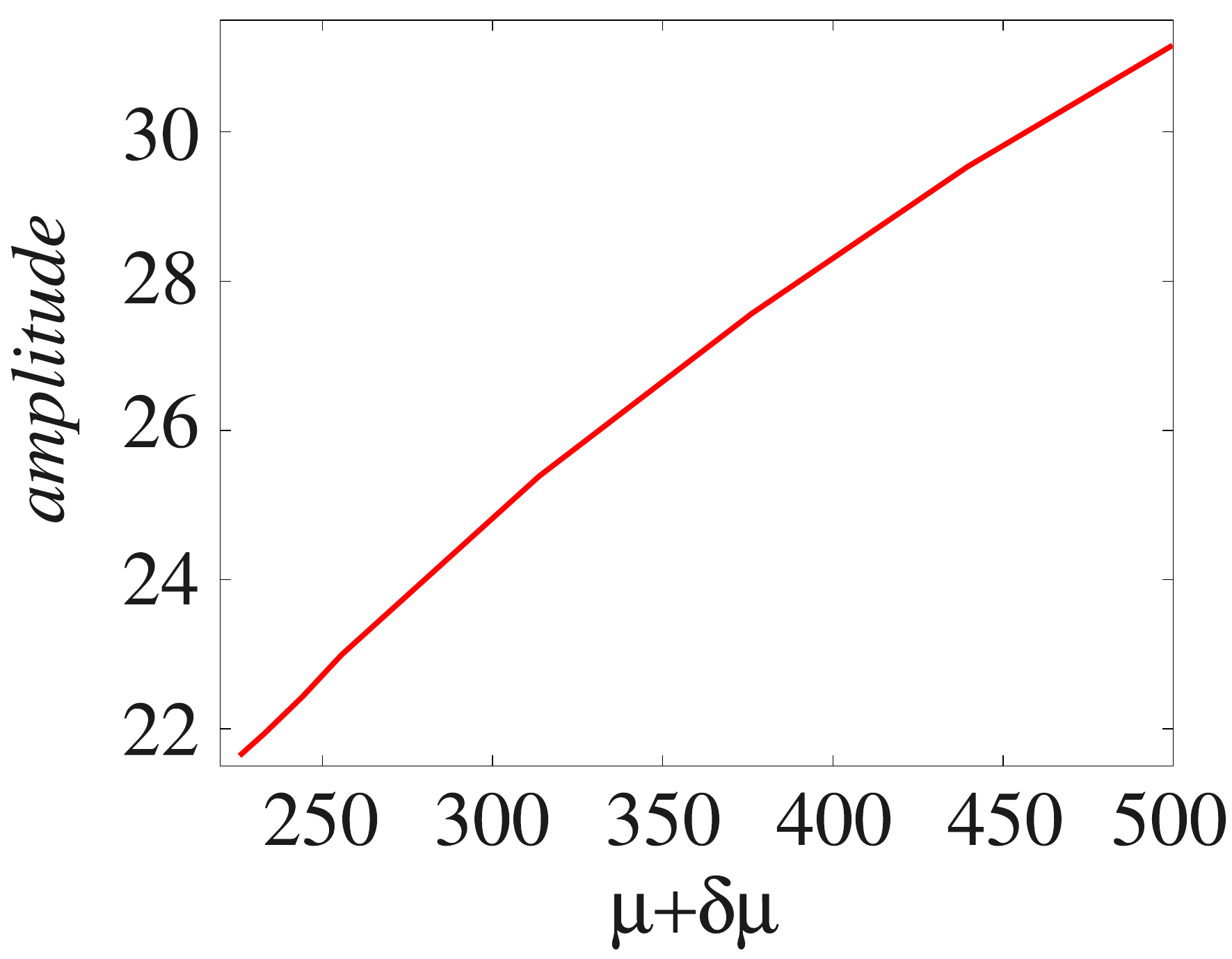}
\caption{(Color online) The same as in Fig. \protect\ref{triangle_curve},
but for the hexagon-shaped vortices with the double topological charge. Only
stable modes are presented in these plots.}
\label{double_curve}
\end{figure}

Figure \ref{double_curve} presents the energy and amplitude of the double
vortices as functions of the effective wavenumber, similar to Fig. \ref%
{triangle_curve} for the triangular vortices with $S=1$. However, only the
stable family of the hexagonal vortices is shown here, as we were not able
to measure characteristics of unstable ones at $\mu <223$. In fact, the
simulations produce no evidence that such unstable modes exist.

\section{Compact triangular vortices}

\begin{figure}[tbp]
\subfigure[]{\includegraphics[width=6cm]{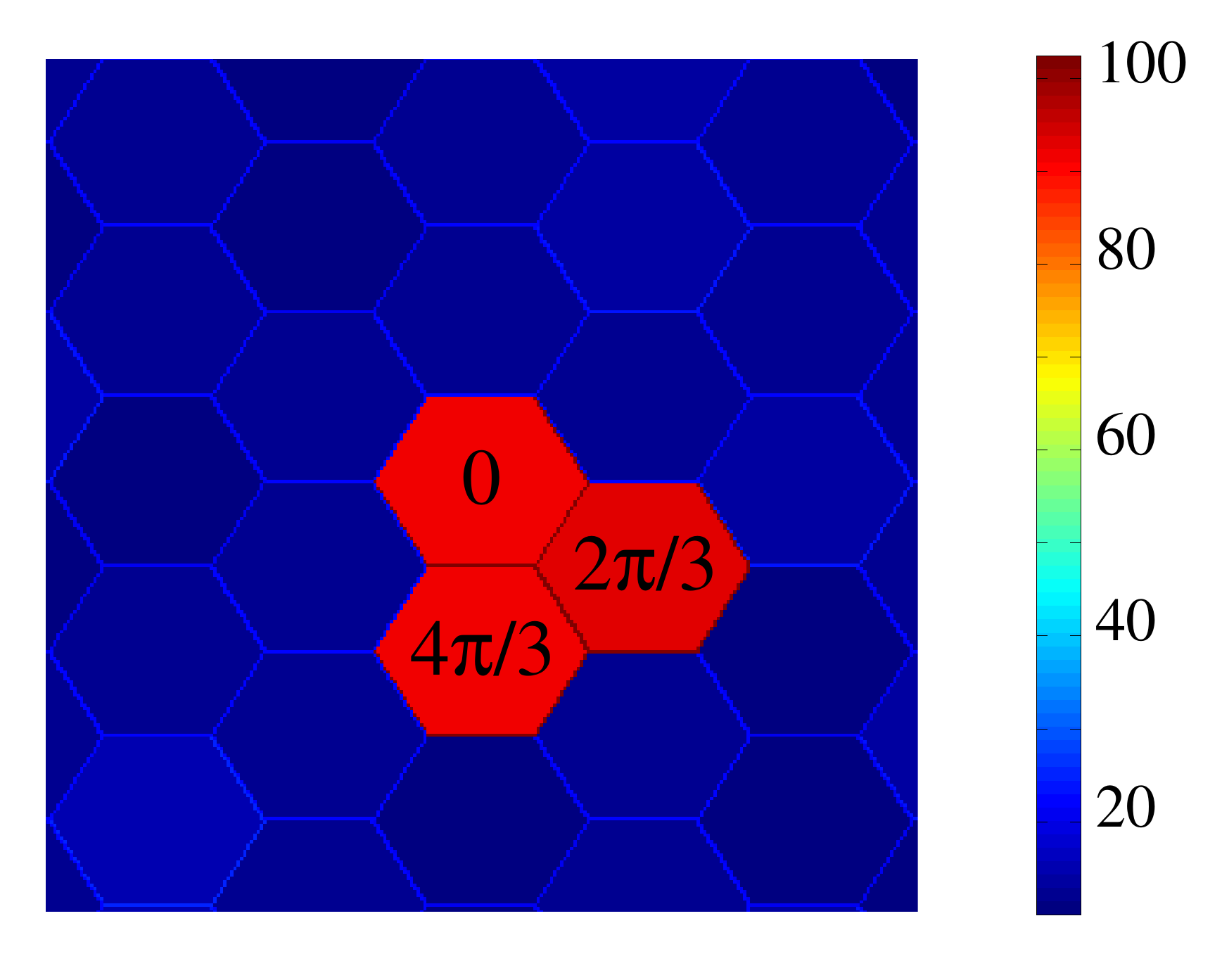}} %
\subfigure[]{\includegraphics[width=6cm]{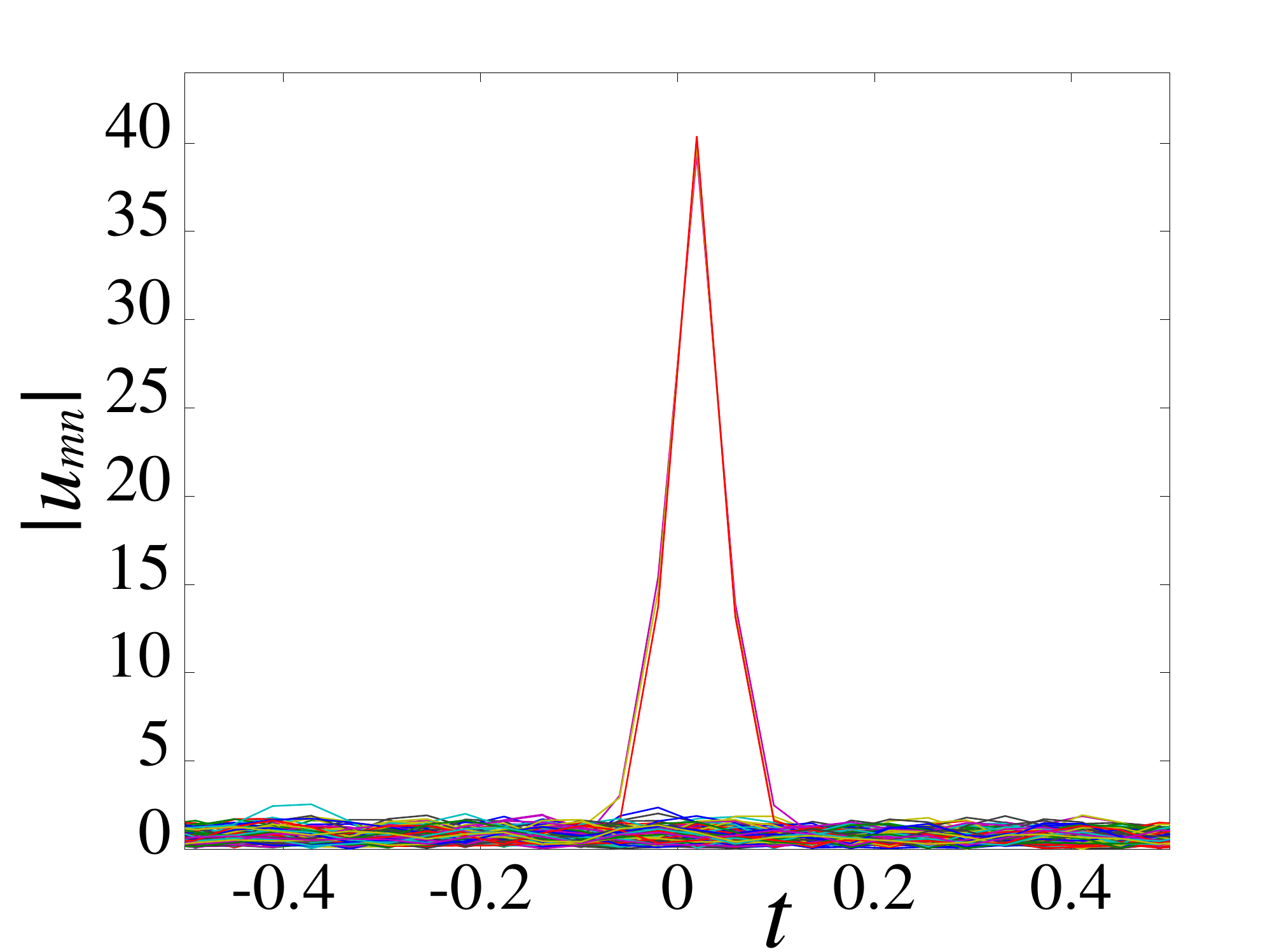}}
\caption{The same as in Fig. \protect\ref{triangle_200}, but for a stable
\textit{compact} triangular vortex with topological charge $S=1$, generated
by the shifted input with $\protect\mu =15$.}
\label{ptttri_15}
\end{figure}

Still another type of stable spatiotemporal patterns can be produced by the
input taken as per Eqs. (\ref{hexag})-(\ref{a}), but centered at an edge of
the original hexagon, i.e., with $m$ and $n$ replaced by $m-2/3$ and $n-1/3$%
, respectively. In this case, the results were collected for $5\leq \mu
\leq 220$
, and, as shown in Fig. \ref{ptttri_15}, the stable structure takes the form
of a vortex with $S=1$, shaped as a \emph{densely packed} triangle, without
an empty site in the center, cf. Fig. \ref{triangle_200}. This structure is
stable for $\mu \geq 11$. It is relevant to stress that this stabilization
threshold is more than an order of magnitude lower than its counterparts for
the triangular and hexagonal spatiotemporal vortices reported in the
previous sections (recall those thresholds were $\mu _{\mathrm{triangle}%
}=182 $ and $\mu _{\mathrm{hexagon}}=223$, respectively). 

In fact, direct simulations initiated by the above-mentioned shifted
input ansatz generate the compact triangle which seems ``noisy". The
noise can be removed by means of the ``temporal filtering", setting
the field equal to zero outside of the main pulse in each core, and
running the additional propagation over $\Delta z=10$. Furthermore,
for $31\leq \mu \leq 60$, direct simulations starting from the
shifted input ansatz lead to a phase instability. For instance, at
$\mu =32$ and $40$, the phases of the three vertices would take
values $0$, $\pi /2$ and $\pi $, instead of those displayed in Fig.
\ref{triangle_200}. Actually, this instability is caused by the fact
that the input is far from the shape of the stable mode, giving rise
to several temporal peaks in each core. If the initial data are
``cleaned up" by nullifying the field outside of the main temporal
pulse, the simulations converge to stable compact triangular
vortices.
%
%
The amplitude of stable compact triangular vortices evolves slowly and almost linearly versus the effective propagation constant $\mu+\delta\mu$: the computed values of the latter range from $\simeq22$ to 78, then the amplitude goes from 40.3 to 41.2.

\section{Collisions between moving vortex solitons}

The availability of stable solutions for the vortex spatiotemporal solitons,
and the obvious Galilean invariance of Eq. (\ref{eq1}) suggest to study
collisions between moving vortices. In particular, it is interesting to
simulate collisions between stable triangular modes shown in Fig. \ref%
{triangle_200}, rotated by angle $\pi /3$ relative to each other, to test a
possibility of their fusion into a full hexagonal vortex of the type
displayed in Fig. \ref{double_223}. This was done taking a pair of the
triangles separated by a relatively large temporal interval, $\Delta t=16$,
for values of $\mu >182$, at which the triangular vortices are stable by
themselves, as shown above. They were set in motion, multiplying them by $%
\exp \left( \pm ik_{0}t\right) $, which, obviously, lends the solitons
velocities $\pm k_{0}$ (in terms of the optical waveguides, these are shifts
of the inverse velocities).

\begin{figure}[tbp]
\includegraphics[width=8cm]{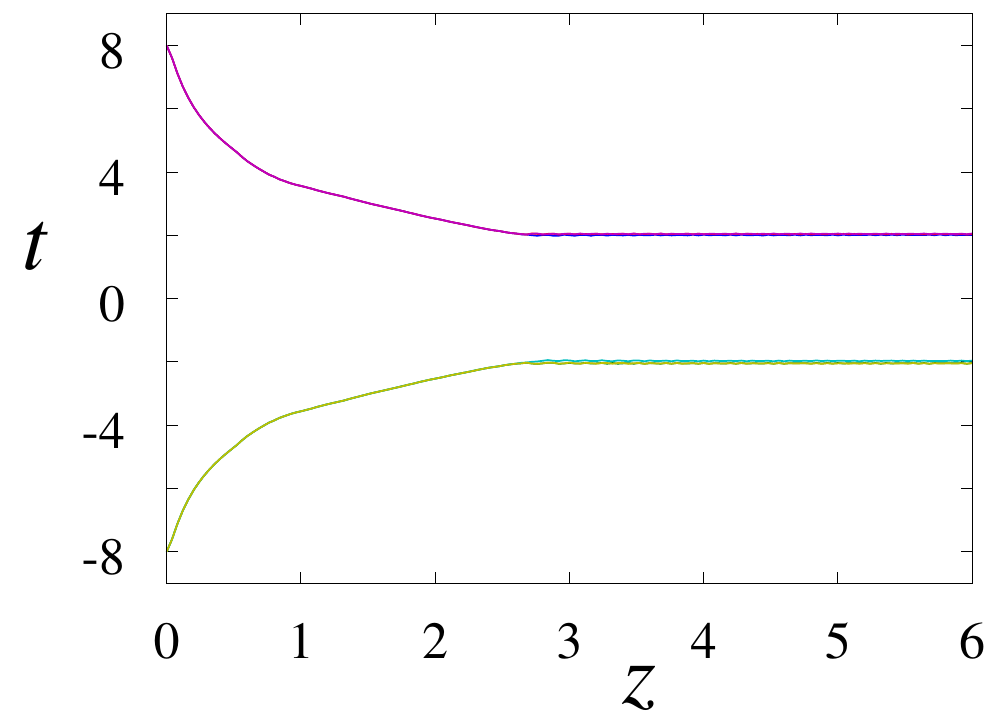}
\caption{Trajectories of the motion of colliding triangular vortices rotated
by angle $\protect\pi /3$ relative to each other, for $\protect\mu =250$ and
velocities $\pm k_{0}=\pm 11$.}
\label{inter11}
\end{figure}

\begin{figure}[tbp]
\includegraphics[width=8cm]{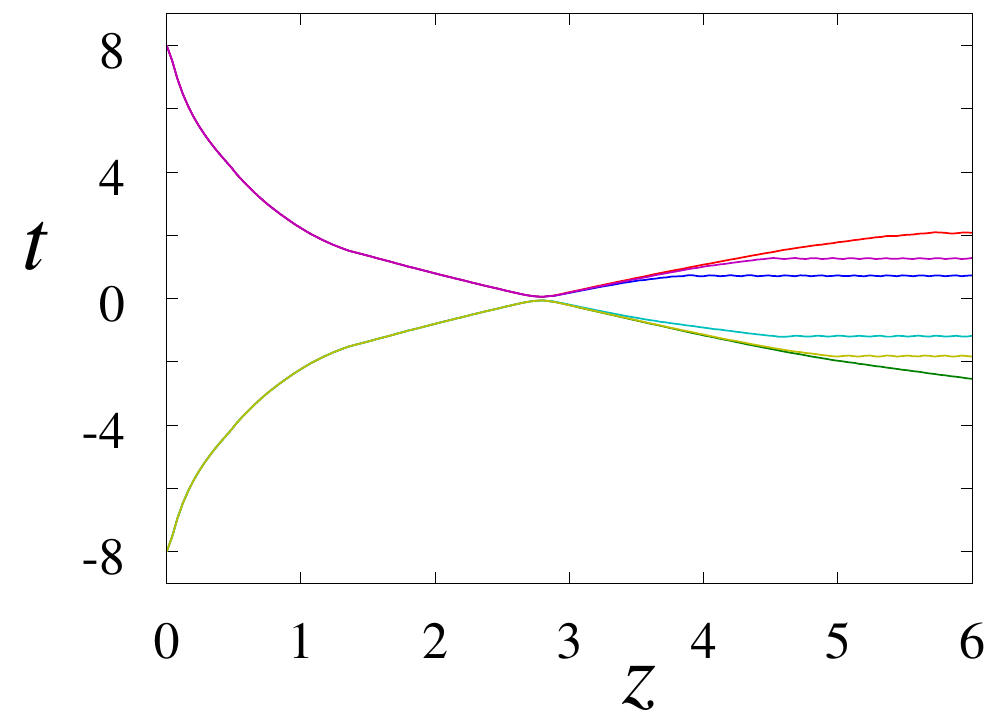}
\caption{The collision of two triangular vortices at $\protect\mu =250$ and $%
k_{0}=12$. Trajectories of individual pulses forming the vortices are
displayed.}
\label{inter12}
\end{figure}

\begin{figure}[tbp!]
\includegraphics[width=8cm]{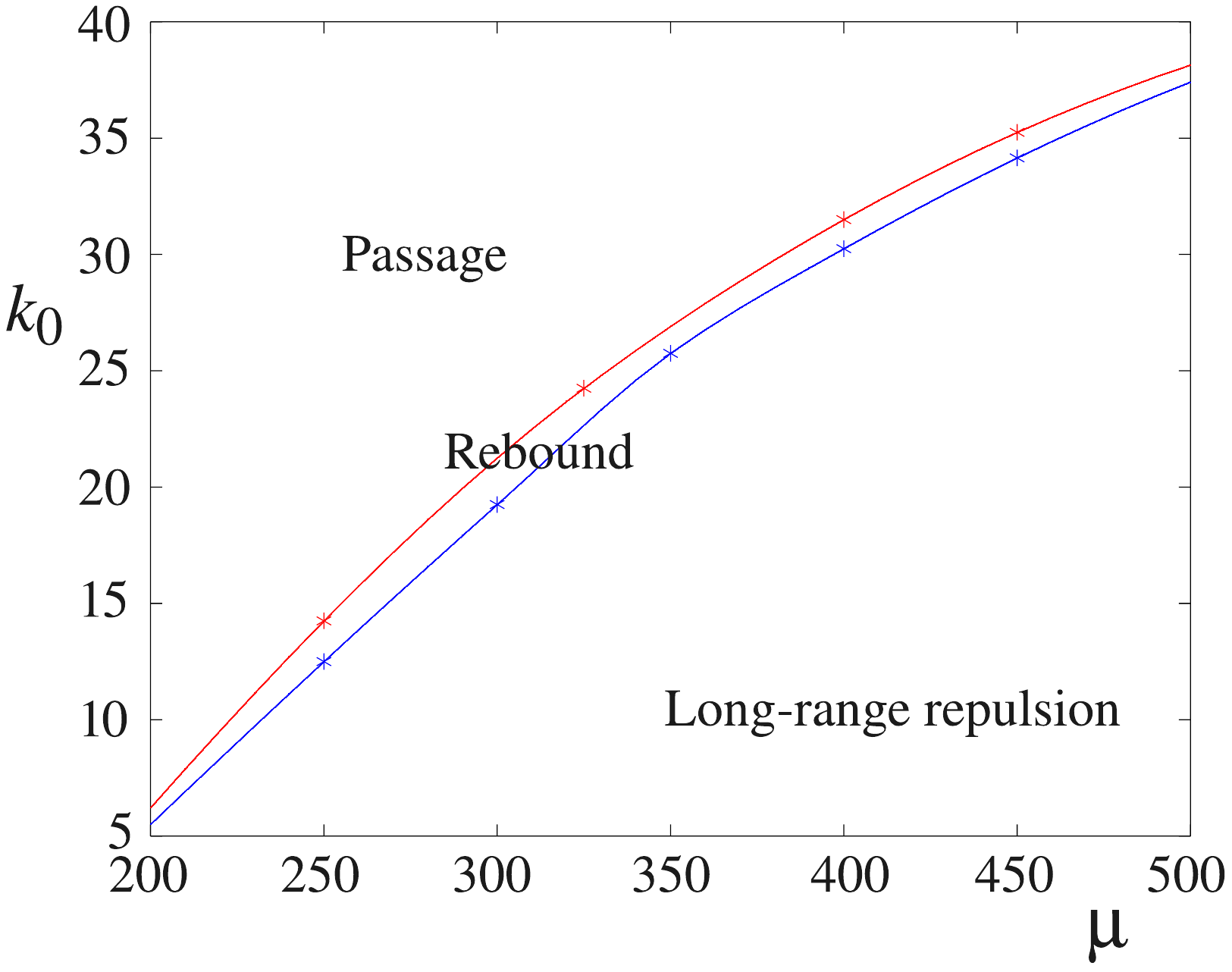}
\caption{(Color online) Regions of different outcomes of collisions between
two mutually symmetric triangular vortices, rotated by angle $\protect\pi /3$.}
\label{li_inter}
\end{figure}

In fact, the fusion of colliding triangles into a hexagon was never
observed. Instead, slowly moving triangles demonstrate a long-range
repulsion and stop at finite distance (but do not bounce back), as shown in
Fig. \ref{inter11}.
At intermediate velocities, the colliding triangular vortices do bounce
back, and eventually they get destroyed by the longitudinal instability
(splitting into uncorrelated temporal pulses in different cores), as shown
in Fig. \ref{inter12}.
At high velocities, the solitons, quite naturally, pass through each other,
loosing some kinetic energy. 
There is a sharp threshold between the rebound regime and the passage. Just
above this threshold, the passing vortices get destroyed by the longitudinal
instability shortly after the collision. Domains corresponding to different
outcomes of the collisions in the $\left( \mu ,k_{0}\right) $ plane are
shown in Fig. \ref{li_inter}.

While the collisions seem elastic with the increase of the velocity, it was
not possible to conclude if the vortices remain stable indefinitely long
after such quasi-elastic collisions. Indeed, since the numerical box has a
finite length, and periodic boundary conditions in $t$ are used, the
moving vortices undergo repeated collisions, loosing some velocity each
time. Eventually, they would be destabilized by a collision occurring at a
lower speed.%

%

\section{Conclusion}

We have introduced a system of parallel waveguides with the linear coupling
between nearest neighbors, based on the hexagonal lattice in the transverse
plane. Each guiding core features the cubic self-attractive nonlinearity.
The system can find straightforward realizations in nonlinear optics, and in
BEC trapped in the corresponding optical lattice. Systematic simulations,
starting with a natural input ansatz for vortical hexagons, reveal three
distinct species of stable semi-discrete spatiotemporal complexes, which are
discrete in the transverse plane and continuous in the longitudinal
direction. These are triangular modes with vorticity $S=1$ and hexagonal
ones with $S=2$, both built with an empty core at the center, and compact
triangles carrying $S=1$, without the central empty core. Collisions between
stable triangular vortices were also studied by means of simulations,
demonstrating the stoppage of the slowly moving vortex solitons,
destabilizing rebounds, and quasi-elastic passage, depending on the
collision velocity.

More complex structure of the arrayed waveguides can be considered as a
generalization of this work (in particular, quasi-periodic lattices). It may
also be interesting to study vortex complexes in two-component models of the
same type.

\section*{Acknowledgements}

This work was supported, in a part, by a grant from the High Council for
Scientific and Technological Cooperation between France and Israel. The work
of DM at Laboratoire de Photonique d'Angers, has been supported by a Senior
Chair Grant from the R\'{e}gion Pays de Loire, France.

\end{document}